\pdfoutput=1

\documentclass[aps,pra,10pt,twocolumn,showpacs,superscriptaddress]{revtex4}
\usepackage{amsmath}
\usepackage{latexsym}
\usepackage{amssymb}
\usepackage{graphics,epstopdf}

\begin{document}

\title{Leggett-type nonlocal realist inequalities without any constraint on the geometrical alignment of measurement settings}

\author{Ashutosh Rai}
\affiliation{S.N.Bose National Center for Basic Sciences, Block JD, Sector III, Salt Lake, Kolkata-700098, India}

\author{Dipankar Home}
\affiliation{Bose Institute, Block - EN, Sector - V, Salt Lake City, Kolkata - 700091, India}

\author{A. S. Majumdar}
\affiliation{S.N.Bose National Center for Basic Sciences, Block JD, Sector III, Salt Lake, Kolkata-700098, India}

\begin{abstract}
Leggett-type nonlocal realist inequalities that have been derived to date are all contingent upon suitable geometrical constraints to be strictly satisfied by the spatial arrangement of the relevant measurement settings. This undesirable restriction is removed in the present work by deriving appropriate forms of nonlocal realist inequalities, one of which involve the least number of settings compared to all such inequalities derived earlier. The way such inequalities would provide a logically firmer basis for a clearer testing of Leggett-type nonlocal realist model vis-a-vis quantum mechanics is explained.
\end{abstract}


\maketitle 
\section{Introduction}
Subsequent to the plethora of studies confirming experimental falsification of Bell-type inequalities \cite{aspect}, thereby ruling out the local realist models in favor of quantum mechanics (QM), the next issue is whether the question of compatibility between QM and its plausible \emph{nonlocal realist models} can be subjected to a deeper scrutiny. To this end, Leggett \cite{leggett} showed an incompatibility between QM and a testable inequality derived for a class of nonlocal realist models which we shall refer to as the Leggett-type nonlocal realist (LNR) model. This, in turn, has motivated a number of theoretical as well as experimental works from different perspectives \cite{groblacher,paterek,branciard0,branciard,colbeck,views,leggett2,eisaman,lee,romero}, including various versions of LNR inequalities. These inequalities involve correlation functions of joint polarization (spin) properties of two spatially separated photons (spin-$\frac{1}{2}$ particles), and have been largely shown to be experimentally violated for the polarization degrees of freedom of photons prepared in a maximally entangled state.

In the initial experiment by Gr\"oblacher \emph{et al.} \cite{groblacher}, though, the form of the LNR inequality that was tested necessitated assuming the invariance of the correlation functions under simultaneous rotation (by the same angle) of the axis of each of the two polarizers. This additional assumption was, however, not required in the subsequent works \cite{paterek,branciard0,branciard} that showed empirical violation of the suitably derived forms of LNR inequalities. Nevertheless, an undesirable feature besets all such studies since different forms of LNR inequalities that have been derived and tested to date hold good only if certain geometrical constraints are \emph{exactly} satisfied by the spatial arrangement of the relevant measurement settings. For example, appropriate to any such inequality, relative orientations of the planes of the relevant measurement settings need to satisfy suitable conditions such as that of orthogonality. Hence, in the experimental tests of these inequalities, even an infinitesimal error in satisfying the required restrictions would make it logically problematic to draw any firm conclusion about the falsification of the LNR model 
 \cite{colbeck}. This loophole is sought to be removed in the present paper by deriving within the general framework of the LNR model two different forms of LNR inequalities that hold good for \emph{any} possible geometrical alignment of the experimental setup. Further, it is important to note that the QM violation of such inequalities can be demonstrated within the experimental threshold visibility already achieved. The other significant feature is that one of our LNR inequalities involves $(3+3)$ number of settings which is the \emph{least} number of settings achieved so far compared to all the LNR inequalities derived earlier.

\section{Leggett's model}
We begin by briefly recapitulating the essence of the LNR model \cite{leggett,leggett2} which regards the whole ensemble of photon pairs emitted from a source to be a disjoint union of subensembles that are assumed to have the following features: (i) In any such subensemble, each pair of photons is characterized by definite values of preassigned polarizations $\hat{u}$ and $\hat{v}$ so that the whole ensemble corresponds to a distribution of values of $\hat{u}$ and $\hat{v}$ denoted by, say, $D(\hat{u},\hat{v})$. (ii) For any given pair belonging to such a subensemble, individual outcomes (denoted by $A$ and $B$) of polarization measurements on
each member of the pair along directions, say, $\hat{a}$ and $\hat{b}$ respectively are assumed to be determined by a hidden variable, say, $\lambda$ whose values are distributed over the pairs comprising the given subensemble with the corresponding distribution function being denoted by $\rho _{(\hat{u},\hat{v})}(\lambda)$. (iii) The outcome of polarization measurement along $\hat{a}$ ($\hat{b}$) for any individual photon in one of the two wings may be non-locally dependent on the choice of the measurement setting pertaining to its spatially separated partner in the other wing, but the statistical result for a given subensemble obtained by averaging over such effects is assumed to satisfy the Malus law. This entails that the relevant mean value depends only on the local setting. Thus, such mean values of outcomes of polarization measurements for the subensembles characterized by $\hat{u}$ and $\hat{v}$ pertaining to the two wings can respectively be written as $\overline{A}(\hat{u})=\int{A(\hat{a},\hat{b},\lambda) \rho_{(\hat{u}, \hat{v})}(\lambda) d\lambda}=\hat{u}\cdot \hat{a}$, and $\overline{B}(\hat{v})=\int{B(\hat{b},\hat{a},\lambda) \rho_{(\hat{u}, \hat{v})}(\lambda) d\lambda}=\hat{v}\cdot \hat{b}$. Then the experimentally observable polarization correlation function for the whole ensemble is expressible as $\langle AB\rangle = \iint \overline{AB}(\hat{u},\hat{v})D(\hat{u}, \hat{v}) d\hat{u} d\hat{v}$ where $\overline{AB}(\hat{u},\hat{v})=\int{A(\hat{a},\hat{b},\lambda)B(\hat{b},\hat{a},\lambda) \rho_{(\hat{u}, \hat{v})}(\lambda) d\lambda}$. 

\section{Derivation of geometrical constraint-free Leggett-type inequalities}
Let us consider that for a pair of emitted photons, $A=\pm 1$ and $B=\pm 1$ are the outcomes observed by two spatially separated partners Alice and Bob performing polarization measurements on each of the photons in the directions $\hat{a}$ and $\hat{b}$ respectively. An outcome $+1$ ($-1$) is associated with a photon getting transmitted (absorbed) through (in) the relevant polarizer. Then, one can easily verify that the algebraic identity $-1+|A+B|=AB=1-|A-B|$ holds true for all the possible outcomes of Alice and Bob. Subsequently, on averaging this relation over any one of the subensembles (characterized by $(\hat{u}, \hat{v})$) mentioned earlier, one obtains $-1+\overline{|A+B|}=\overline{AB}=1-\overline{|A-B|}$, where the bar notation denotes averaging over the hidden variables within the given subensemble. Since the average of the modulus is greater or equal to the modulus of the averages, therefore, at the level of subensembles one gets, $-1+|\overline{A}+\overline{B}|\leq \overline{AB}\leq 1-|\overline{A}-\overline{B}|$, which can be rewritten as the following inequality
\begin{equation}
 |\overline{A}\pm \overline{B}|\leq 1\pm \overline{AB}.\label{eqn0} 
 \end{equation}
 
Next, $\grave{a}$ la Branciard \emph{et al.} \cite{branciard}, consider one measurement setting $\hat{a}$, with the corresponding outcome $A$, for Alice, and two measurement settings $\hat{b},\hat{b'}$, with the corresponding outcomes $B, B'$ for Bob. Applying the inequality (\ref{eqn0}) for the sets $\{A,B\}$ and $\{A,B'\}$ respectively, together with the use of the triangle inequality, one can obtain the following inequality $|\overline{AB}\pm \overline{AB'}|\leq 2-|\overline{B}\mp \overline{B'}|$. Then, by invoking the Malus law on the right hand side of the preceding inequality and averaging over the distribution $D(\hat{u},\hat{v})$, one gets
\begin{eqnarray}
|\langle AB\rangle +\langle AB'\rangle|\leq 2-\iint|\hat{v}\cdot (\hat{b}-\hat{b}')|D(\hat{u},\hat{v})d\hat{u}d\hat{v} \label{eqn1}\\
|\langle AB\rangle -\langle AB'\rangle|\leq 2-\iint|\hat{v}\cdot (\hat{b}+\hat{b}')|D(\hat{u},\hat{v})d\hat{u}d\hat{v} \label{eqn2}
\end{eqnarray}
Now, at this stage, comes the crucial ingredient of our derivation by considering two different categories of settings that would enable us to derive the desired forms of the LNR inequalities. Note that in this derivation there is no geometrical restriction on the spatial arrangement, once any particular type of combination of settings is specified.

\subsection{Category I settings} 
Category I comprising of suitable combinations of measurement settings used for deriving our first LNR inequality, pertains to the inequality (\ref{eqn1}). Here we consider the combinations of settings $\{(\hat{a}_{i},\hat{b}_{i}), (\hat{a}_{i}, \hat{b'}_{i})\}$ where $i\in \{1,2,3\}$ and, say, $\beta _{i}\in (-\pi,\pi)$ is the angle between the pair $(\hat{b}_{i},\hat{b'}_{i})$. Let $\hat{b} _{i}-\hat{b'}_{i}=2 \sin (\frac{\beta _{i}}{2})\hat{n}_{i}$ where the unit vectors $\hat{n}_{i}$'s are \emph{linearly independent}. Then, from (\ref{eqn1}), after adding the corresponding inequalities for the combinations of settings $\{(\hat{a}_{i},\hat{b}_{i}), (\hat{a}_{i}, \hat{b'}_{i})\}$, it follows that
 \begin{eqnarray}
 \frac{1}{3}\sum _{i}|\langle A_{i}B_{i}\rangle +\langle A_{i}B_{i}'\rangle| \leq \hspace{100pt}\nonumber \\ 2-\frac{2}{3} \sin (\frac{\beta_{*}}{2})
 \iint F_{n}(\hat{v})D(\hat{u},\hat{v})d\hat{u}d\hat{v} \label{eqn3}
 \end{eqnarray}
 where, $\beta_{*}=\mbox{min}\{|\beta _{1}|,|\beta _{2}|,|\beta _{3}|\}$ and $F_{n}(\hat{v})=\sum _{i}|\hat{v}\cdot \hat{n}_{i}|$.
 
\subsection{Category II settings} 
Category II comprising appropriate combinations of measurement settings involved in our second LNR inequality, pertains to the inequality (\ref{eqn2}). Here we consider the combinations of settings $\{(\hat{a}_{i}, \hat{b}_{i}),(\hat{a}_{i}, \hat{b}_{i\oplus 1})\}$ where $\oplus$ represents addition modulo $3$, and $i\in \{1,2,3\}$. Let $\delta _{i}\in (-\pi , \pi)$ be the angle between the pair $(\hat{b}_{i}, \hat{b}_{i\oplus 1})$, whence $\hat{b} _{i}+\hat{b}_{i\oplus 1}=2 \cos (\frac{\delta _{i}}{2})\hat{m}_{i}$ where $\hat{m}_{i}$'s represent three \emph{linearly independent} unit vectors. Then, from (\ref{eqn2}), after adding the corresponding inequalities for the combination of settings $\{(\hat{a}_{i}, \hat{b}_{i}),(\hat{a}_{i}, \hat{b}_{i\oplus 1})\}$, it follows that
\begin{eqnarray}
 \frac{1}{3}\sum _{i}|\langle A_{i}B_{i}\rangle -\langle A_{i}B_{i\oplus 1}\rangle| \leq \hspace{100pt}\nonumber \\2-\frac{2}{3} \cos (\frac{\delta^{*}}{2})
 \iint \!\!F_{m}(\hat{v})D(\hat{u},\hat{v})d\hat{u}d\hat{v} \label{eqn4}
 \end{eqnarray}
 where $\delta^{*}=\mbox{max}\{|\delta _{1}|,|\delta _{2}|,|\delta _{3}|\}$ and $F_{m}(\hat{v})=\sum _{i}|\hat{v}\cdot \hat{m}_{i}|$. 
 
\subsection{Lower bound for the functions $F_n(\hat{v})$ and $F_m(\hat{v})$}
Note that the right hand sides of the inequalities (\ref{eqn3}) and (\ref{eqn4}) still involve the unobservable supplementary variables $\hat{u}$ and $\hat{v}$. Thus, in order to recast them in experimentally verifiable forms, we need to derive the respective \emph{lower bounds}, say $L_n$ and $L_m$, for the functions $F_{n}(\hat{v})$ and $F_{m}(\hat{v})$. These lower bounds are obtained by using the following \emph{Theorem}.

\textbf{Theorem}: On the Poincar{\'e} sphere, given three linearly independent unit vectors $\hat{e}_{1}$, $\hat{e}_{2}$, $\hat{e}_{3}$ and a variable unit vector $\hat{v}$, the minimum value, say $L$, of the function $ F(\hat{v})= |\hat{e}_{1}\cdot \hat{v}|+ |\hat{e}_{2}\cdot \hat{v}|+ |\hat{e}_{3}\cdot \hat{v}|$ is given by the formula $L = \frac{|\hat{e}_{1}\cdot (\hat{e}_{2}\times \hat{e}_{3})|}{\mbox{max}\{ |\hat{e}_{1}\times \hat{e}_{2}|, |\hat{e}_{2}\times \hat{e}_{3}|, |\hat{e}_{3}\times \hat{e}_{1}| \}}$.

\begin{figure}[!ht]
\resizebox{6cm}{6cm}{\includegraphics{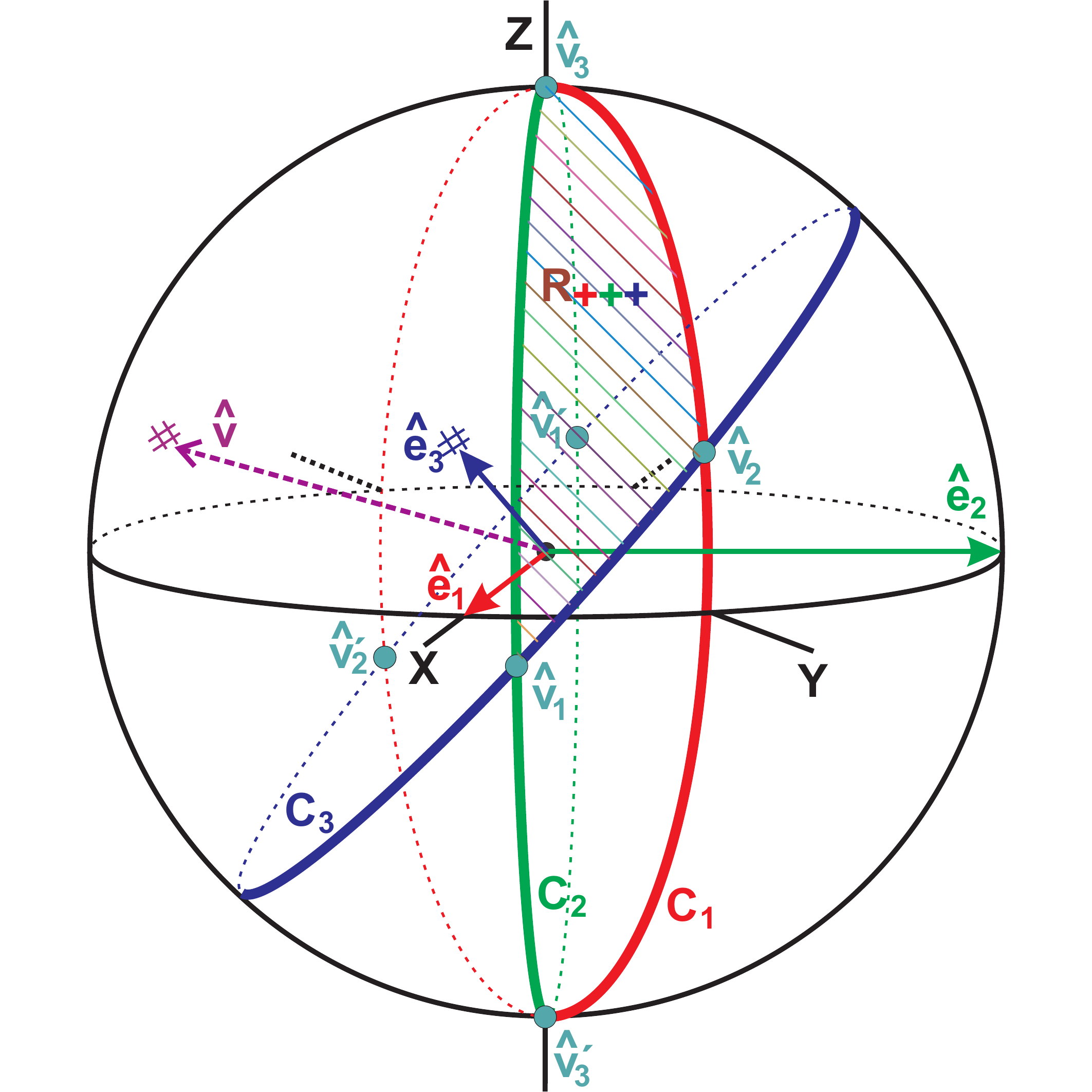}}
\caption{ (Color online) On the Poincar{\'e} sphere, $\hat{e}_{i}$'s for $i \in \{1,2,3\}$ are three linearly independent unit vectors and $\hat{v}$ is a variable unit vector. Three great circles $C_{i}$'s lie in respective planes orthogonal to $\hat{e}_{i}$'s. The intersection points of two great circles $C_{i\oplus 1}$ and $C_{i\oplus 2}$ are denoted by $\hat{v}_{i}$ and $\hat{v}_{i'}$. The triangular region $R_{+++}$ with vertices $\hat{v}_{1}$, $\hat{v}_{2}$, $\hat{v}_{3}$ is defined by relations $\hat{e}_{1}\cdot \hat{v}\geq 0$, $\hat{e}_{2}\cdot \hat{v}\geq 0$, and $\hat{e}_{3}\cdot \hat{v}\geq 0 $. \label{fig2}}
\end{figure}

\textbf{Proof}: The minimum value of $F(\hat{v})$ would not depend on the choice of the coordinate axes. Thus, for convenience, let the $X$ axis lie along $\hat{e}_{1}$ and the $XY$ plane contain $\hat{e}_{2}$. Therefore, according to our choice, $\hat{e}_{i}$'s can be represented as follows: $\hat{e}_{1}=(1,0,0)$, $\hat{e}_{2}=(b_{1},b_{2},0)$, $\hat{e}_{3}=(c_{1},c_{2},c_{3})$ where $b_{2},c_{3}\neq 0$. Further, we observe that the three great circles $C_{i}$, defined by $\hat{e}_{i}\cdot \hat{v}=0$, divide the surface of the Poincar{\'e} sphere into $8$ non-overlapping (except on the boundaries) regions, $R_{\xi_{1}\xi_{2}\xi_{3}}$, defined by the constraints $\xi_{1}\hat{e}_{1}\cdot \hat{v}\geq 0$, $\xi_{2}\hat{e}_{2}\cdot \hat{v}\geq 0$, and $\xi_{3}\hat{e}_{3}\cdot \hat{v}\geq 0$, where $\xi_{1},\xi_{2},\xi_{3} \in \{+,-\}$ [see Fig. \ref{fig2}].

Let us first minimize the function $F(\hat{v})$ in any one of the restricted regions, say, $R_{+++}$ where it takes the form, $F(\hat{v}_{+++})=\hat{e}_{1}\cdot \hat{v}+ \hat{e}_{2}\cdot \hat{v}+ \hat{e}_{3}\cdot \hat{v}$ (here $\hat{v}_{+++}$ denote vectors belonging to the region $R_{+++}$).
We first show that $F({\hat{v}_{+++}})$ cannot attain the minimum at some interior point of $R_{+++}$. Note that, since $F({\hat{v}_{+++}})$ is a smooth function, showing that there is no stationary point of local minimum in the interior of $R_{+++}$ would be sufficient. For this, let us consider a function $f({\hat{v}})=\hat{e}_{1}\cdot \hat{v}+ \hat{e}_{2}\cdot \hat{v}+ \hat{e}_{3}\cdot \hat{v}$ defined over all the points of the Poincar{\'e} sphere. Note that, $f(\hat{v})= F(\hat{v})$ for any $\hat{v}\in R_{+++}$. Let $\hat{v}=(\sin \theta \cos \phi ,\sin \theta \sin \phi ,\cos \theta)$ with $0\leq \theta \leq \pi$ and $-\pi< \phi \leq \pi$. Then, $f({\hat{v}})= f(\theta ,\phi)=p\sin \theta \cos \phi +q\sin \theta \sin \phi +r\cos \theta$ where $p=1+b_{1}+c_{1}$, $q=b_{2}+c_{2}$, and $r=c_{3}$. At the stationary points of $f(\theta ,\phi)$, $\partial_{\phi}f= \sin \theta (-p\sin \phi +q\cos \phi)=0 $ and $\partial_{\theta}f=\cos \theta (p\cos \phi +q\sin \phi)-r\sin \theta =0$. However, among such stationary points, the point belonging to the interior of the region $R_{+++}$ would satisfy 
$(\partial_{\phi\phi}f)(\partial_{\theta\theta}f)-\partial_{\phi\theta}f > 0$ and $\partial_{\theta\theta}f<0$, which is the condition of maximum. Thus, $F(\hat{v}_{+++})$ can attain its minimum value only on some boundary point of the region $R_{+++}$.

Next, we find that the minimum value of $F(\hat{v}_{+++})$ is actually attained at any one or more vertices of the triangular region $R_{+++}$; these vertices are given by $\hat{v}_{1}=\mbox{sgn}(b_{2}c_{3})\frac{\hat{e}_{2}\times \hat{e}_{3}}{|\hat{e}_{2}\times \hat{e}_{3}|}$, $\hat{v}_{2}=\mbox{sgn}(b_{2}c_{3})\frac{\hat{e}_{3}\times \hat{e}_{1}}{|\hat{e}_{3}\times \hat{e}_{1}|}$, $\hat{v}_{3}=\mbox{sgn}(c_{3})\frac{\hat{e}_{1}\times \hat{e}_{2}}{|\hat{e}_{1}\times \hat{e}_{2}|}$ where $\mbox{sgn}(z)=+1(-1)$ for $z>0$($z<0$). Here, first note that, the intersection of $C_{i}$ with $R_{+++}$ defines a side of the triangle $R_{+++}$. Now, if we restrict the domain of $f(\hat{v})$ on a great circle $C_{i}$, then it can be shown that, for any $i$, there is no stationary point of minimum of $f(\hat{v})$ in the interior of the corresponding side of the triangle $R_{+++}$ (see Appendix A). Hence, now we can conclude that the minimum value of $F(\hat{v}_{+++})$ is attained only at some vertices of the region $R_{+++}$.

 Note that the above proven result is true for any arbitrarily specified set of linearly independent unit vectors $\{\hat{e}_1, \hat{e}_2, \hat{e}_3\}$ i.e., if one chooses, say, some other set $\{\hat{e}^*_1, \hat{e}^*_2, \hat{e}^*_3\}$ then the minimum value of the corresponding function $ F^*(\hat{v})= |\hat{e}^*_{1}\cdot \hat{v}|+ |\hat{e}^*_{2}\cdot \hat{v}|+ |\hat{e}^*_{3}\cdot \hat{v}|$ in a suitably defined region $R^*_{+++}$ is attained at one or more of its vertices.

Finally, with the help of the above shown result and exploiting the symmetries of the function $F(\hat{v})$ we show that the desired minimum value is $\mbox{min}[F(v_{+++})]$. For this, let us consider a repartitioning of the set of points on the Poincar{\'e} sphere defined by $R^{\chi_1\chi_2\chi_3}_{\xi_1\xi_2\xi_3}=\{\hat{v}: \xi_i(\chi_i\hat{e}_i)\cdot \hat{v}\geq 0, ~\forall i\in \{1,2,3\}\}$ for some fixed  $\chi_1, \chi_2, \chi_3 \in \{+,-\}$ (Observe that there are $8$ such ways of partitioning and in the new notation the partition represented by $R_{\xi_1\xi_2\xi_3}\equiv R^{+++}_{\xi_1\xi_2\xi_3}$). Then, we note the following two features for above type of repartition: (i) The relevant function $F^{\chi_1\chi_2\chi_3}(\hat{v})=|(\chi_1\hat{e}_{1})\cdot \hat{v}|+ |(\chi_2\hat{e}_{2})\cdot \hat{v}|+ |(\chi_3\hat{e}_{3})\cdot \hat{v}|$ remains invariant for any choice of $\chi_i$'s $\in \{+,-\}$ and (ii) $ R_{\chi_1\chi_2\chi_3}\equiv R^{+++}_{\chi_1\chi_2\chi_3}\cong R^{\chi_1\chi_2\chi_3}_{+++}$. Since earlier we have shown that for any partition the minimum value of $F^*(\hat{v}_{+++})$ in the corresponding region $R^*_{+++}$ can only be  attained at one or more of its vertices, applying the property (i) and (ii) we can now conclude that minimum value of $F(\hat{v})$ in a region $R_{\xi_1 \xi_2 \xi_3}$ for any $\xi_1$, $\xi_2$, $\xi_3$ $\in \{+,-\}$ is attained at one or more of its vertices. Thus, the required global minimum value $L$ is attained at some point(s) belonging to the set $\{\pm\hat{v}_1, \pm\hat{v}_2, \pm\hat{v}_3\}$. Now, the use of the symmetry $F(-\hat{v}) = F(\hat{v})$ gives $L= \mbox{min}\{F(\hat{v}_{1}),F(\hat{v}_{2}),F(\hat{v}_{3})\}=\frac{|\hat{e}_{1}\cdot (\hat{e}_{2}\times \hat{e}_{3})|}{\mbox{max}\{ |\hat{e}_{1}\times \hat{e}_{2}|, |\hat{e}_{2}\times \hat{e}_{3}|, |\hat{e}_{3}\times \hat{e}_{1}| \}}$.\hspace{0 mm}$\Box$

\subsection{Two testable forms of LNR inequalities}
By applying the above proven theorem, together with the use of the normalization relation $\int\int D(\hat{u},\hat{v})d\hat{u}d\hat{v}=1$, to the inequalities (\ref{eqn3}) and (\ref{eqn4}), we obtain respectively the following two forms of experimentally testable LNR inequalities
\begin{equation}
\frac{1}{3}\sum _{i}|\langle A_{i}B_{i}\rangle +\langle A_{i}B_{i}'\rangle|\leq 2-\frac{2}{3} \sin (\frac{\beta_{*}}{2}) \times L_{n}\label{eqn6}
\end{equation}
\begin{equation}
\frac{1}{3}\sum _{i}|\langle A_{i}B_{i}\rangle -\langle A_{i}B_{i\oplus 1}\rangle|\leq 2-\frac{2}{3} \cos (\frac{\delta^{*}}{2})\times L_{m}\label{eqn7}
\end{equation}
For these two experimentally testable forms of the LNR inequalities, respective lower bounds $L_{n,m}$ for functions $F_{n,m}(\hat{v})= |\hat{e}_{1}\cdot \hat{v}|+ |\hat{e}_{2}\cdot \hat{v}|+ |\hat{e}_{3}\cdot \hat{v}|$ with $\hat{e}_{i}$'s corresponding to $\hat{n}_{i}$'s or $\hat{m}_{i}$'s for $F_{n}$ or $F_{m}$ respectively, can be equivalently expressed by the following convenient expression (see Appendix B for a proof)

\begin{equation}
L_{n,m}(\alpha _{12},\alpha _{23},\alpha _{31})=\frac{\Biggl(1\!\!-\!\!\!\displaystyle\sum_{\substack{1\leq i\leq 3, \\j=i\oplus 1}}\!\!\!\cos ^{2}\alpha _{ij} + 2\!\!\!\!\displaystyle\prod_{\substack{1\leq i\leq 3, \\j=i\oplus 1}}\!\!\!\cos \alpha _{ij}\Biggr)^{\frac{1}{2}}}{\mbox{max}\{\sin \alpha _{12} ,\sin \alpha _{23} ,\sin \alpha _{31}\}} \label{eqn5}
\end{equation}
where $\alpha _{ij}\in (0,\pi)$ denotes the angle between a pair of vectors $\{\hat{e}_{i}, \hat{e}_{j}\}$ for $i,j \in \{1,2,3\}$.

\section{Salient features of the LNR inequalities (\ref{eqn6}) and (\ref{eqn7})}
First, let us focus on the LNR inequality (\ref{eqn6}). Given the way this inequality has been derived by us, Alice and Bob are both free to arbitrarily choose their measurement settings, and given their choices, the \emph{LNR bound} (the right hand side) on the combination of correlation functions (the left hand side) can be calculated with help of the formula (\ref{eqn5}).

Note that the inequality derived and experimentally tested by Branciard \emph{et al.} \cite{branciard} is a special case of the inequality (\ref{eqn6}) by assuming a specific geometrical constraint that requires $\hat{n}_{i}$'s to be mutually orthogonal and $|\beta_{1}|=|\beta_{2}|=|\beta_{3}|$. For a photon pair prepared in a pure singlet state, one can show that settings for observing the maximum violation of the inequality (\ref{eqn6}) are in which Bob's choices are such that the three directions $\hat{n}_{i}$'s of $(\hat{b}_{i}-\hat{b'}_{i})$ are orthogonal where $|\beta_{1}|=|\beta_{2}|=|\beta_{3}|\approx 36.9^o$, with $\hat{a}_{i}$'s chosen by Alice to be along the directions of $\hat{b}_{i}+\hat{b'}_{i}$ (see Appendix B for a proof). 
While the magnitude of the maximum possible violation of the inequality (\ref{eqn6}) corresponds to the threshold visibility of $94.3$\%.

\begin{figure}[!ht]
\resizebox{6.5cm}{7.5cm}{\includegraphics{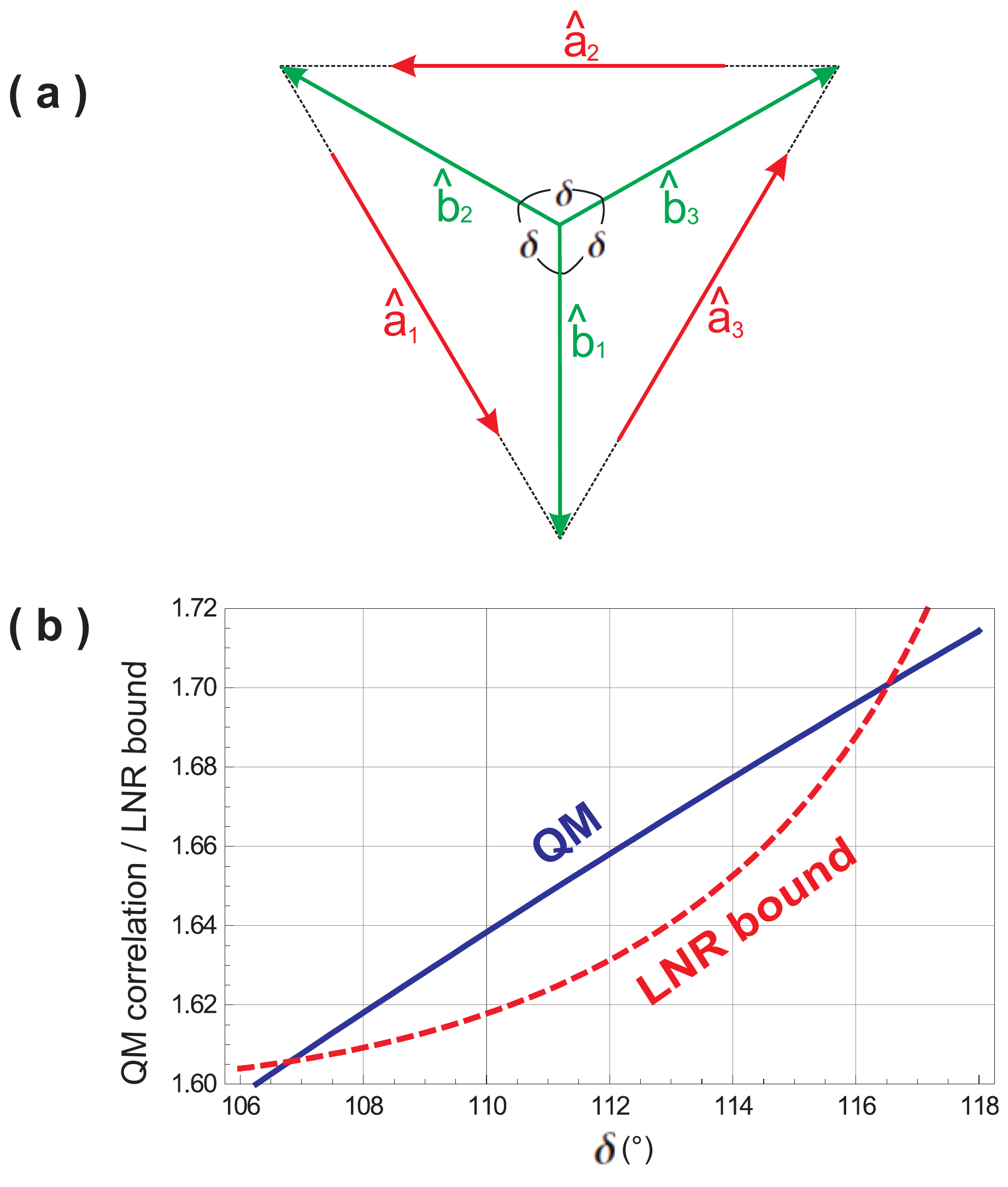}}
\caption
{(Color online) \textbf{(a)} As an illustrative example, a class of symmetric configurations for observing QM violation of LNR inequality (\ref{eqn7}) for a pure singlet state is shown. For Bob's measurement settings $\hat{b}_{i}$'s, where $i \in \{1,2,3\}$, the angle between any pair of settings is $\delta$. Alice's measurement settings $\hat{a}_{i}$'s are along the directions $\hat{b}_{i}-\hat{b}_{i\oplus 1}$\label{fig1}. 
\textbf{(b)} The dotted line shows LNR upper bounds and the bold line shows corresponding QM values of the left hand side of the LNR inequality (\ref{eqn7}) as $\delta$ is varied. A range of QM violations of the inequality (\ref{eqn7}) is obtained for $\delta \in [106.8^o,116.5^o]$, with the maximum violation occurring at $\delta \approx 112.63^o$.}
\end{figure}

Now, considering the LNR inequality (\ref{eqn7}), we note that for a pure singlet state, violations of (\ref{eqn7}) by the QM predictions can be shown, for example, by taking a class of symmetric configurations in which the angles between Bob's measurement settings satisfy $\delta_{1}=\delta_{2}=\delta_{3}=\delta$ and Alice's measurement settings $\hat{a}_{i}$'s are along the directions $\hat{b}_{i}-\hat{b}_{i\oplus 1}$ [see Fig.\ref{fig1}(a)]. For such configurations, QM violations of the inequality (\ref{eqn7}) occur within the domain $\delta \in [106.8^o,116.5^o]$ where the maximum violation is obtained for $\delta \approx 112.63^o$ [see Fig.\ref{fig1}(b)]. In this case, the maximum value of the ratio of the right hand bound and the corresponding QM value of the left hand side is given by $0.9836$, meaning that the threshold visibility in the relevant experiment that is required to show the QM violation of the LNR inequality (\ref{eqn7}) is $98.36 \%$. Now, note that already the visibility above $98.4\%$ was achieved in an experiment by Paterek \emph{et al.} \cite{paterek} where a LNR inequality involving $7$ measurement settings for Bob and $3$ for Alice was tested. A similar work \cite{branciard0} was also reported using a family of LNR inequalities involving $2N$ and $4N$ ($N\geq2$) number of settings for Alice and Bob respectively. 

Thus, an additional significance of our LNR inequality (\ref{eqn7}) lies in involving \emph{lesser} number, only $3$ measurement settings for Bob and $3$ for Alice such that the threshold visibility required to test this inequality is experimentally realizable. An open problem nevertheless remains whether a testable incompatibility can be shown between QM and the LNR model by using even lesser number of settings in either wing without, of course, taking recourse to any additional assumption like that of rotational invariance of correlation functions.

\section{Concluding remarks.}
A generic property of the LNR inequalities is that while the left hand side of any such inequality involves experimentally measurable quantities (correlation functions), the LNR bound (the right hand side) of such an inequality, \emph{unlike} any Bell-type inequality, is not just a number fixed by the general assumptions used in the relevant derivation; instead, it depends on the \emph{choice} of the geometrical configuration of measurement settings. Thus, for different configurations of settings, say, $\emph{S}_{1}$, $\emph{S}_{2}$, $\emph{S}_{3}$... there are corresponding LNR bounds $\emph{B}_{1}$, $\emph{B}_{2}$, $\emph{B}_{3}$.... Therefore, if due to experimental imprecision, the actual settings deviate from the required configuration within a certain domain that can be estimated by the experimenter, there will be a corresponding range of LNR bounds. As a consequence, the experimental violation of any relevant LNR inequality can be unambiguously concluded only if the supremum of such a range of LNR bounds is violated. However, an estimation of such a range of LNR bounds by taking into account all possible imprecisions that may occur in realizing the required configuration of settings is severely restricted for any of the LNR inequalities derived earlier \cite{leggett,groblacher,paterek,branciard0,branciard}. This is essentially because the validity of any such inequality is in itself contingent upon certain geometrical constraints being strictly satisfied by the measurement settings. 
Herein lies the central significance of our LNR inequalities (\ref{eqn6}) and (\ref{eqn7}) in enabling a more logically conclusive test of the LNR model vis-a-vis QM than that has been hitherto possible---this is because the forms of the LNR inequalities derived in the present paper are \emph{free} from any constraint on the spatial alignments of the relevant measurement settings.

\appendix
\section{Discussion of details in the proof of the minimum value of $F(\hat{v}_{+++})$ in the triangular region $R_{+++}$}
Recall that an expression for the function $f({\hat{v}})=\hat{e}_{1}\cdot \hat{v}+ \hat{e}_{2}\cdot \hat{v}+ \hat{e}_{3}\cdot \hat{v}$ defined over the points of the Poincar{\'e} sphere in terms of $(\theta,\phi)$ coordinates ($0\leq\theta\leq\pi$ and $-\pi<\phi\leq \pi$), where $p=1+b_{1}+c_{1}$, $q=b_{2}+c_{2}$ and $r=c_{3}$, is
\begin{equation}
f(\theta ,\phi)=(p\cos \phi +q \sin \phi)\sin \theta +r\cos \theta
\end{equation}
Now, analyzing this expression, we in detail show the following 
\subsection{The minimum value of $F(\hat{v}_{+++})$ is not attained at any interior point of the region $R_{+++}$ }
At the stationary points of $f(\theta ,\phi)$ we have
\begin{eqnarray}
\partial_{\phi}f&=& \sin \theta (-p\sin \phi +q\cos \phi)=0 \\
\partial_{\theta}f&=&\cos \theta (p\cos \phi +q\sin \phi)-r\sin \theta =0
\end{eqnarray}
Therefore, at some stationary point, say $(\theta_{0},\phi_{0})$, in the interior of $R_{+++}$, since $\sin\theta_{0}\neq0$, the following equations must be satisfied
\begin{eqnarray}
-p\sin \phi_{0} +q\cos \phi_{0}&=&0 \\
\cos \theta_{0} (p\cos \phi_{0} +q\sin \phi_{0})&=& r\sin \theta_{0} 
\end{eqnarray}
Then, at $(\theta_{0},\phi_{0})$ we obtain
\begin{widetext}
\begin{eqnarray}
[(\partial_{\phi\phi}f)(\partial_{\theta\theta}f)-\partial_{\phi\theta}f]_{(\theta_{0},\phi_{0})}&=&\{(p\cos\phi_{0}+q\sin\phi_{0})^{2}+r^{2}\}\sin^{2}\theta_{0}>0 \hspace{5mm}(\mbox{since}\hspace{2mm} r=c_{3}\neq0)\\
 (\partial_{\theta\theta}f)_{(\theta_{0},\phi_{0})}&=&-f(\theta_{0},\phi_{0})<0
\end{eqnarray}
\end{widetext}
Thus, given a stationary point $(\theta_{0},\phi_{0})$, it must be a point of maximum. Consequently, the minimum cannot lie in the interior of the region $R_{+++}$.
\subsection{The minimum value of $F(\hat{v}_{+++})$ is not attained at any interior point on the sides of the triangular region $R_{+++}$}
Let us first consider the side of the triangle $R_{+++}$ on the corresponding great circle $C_{1}$. Note that the interior points of this side is defined by
\begin{eqnarray}
\hat{e}_{1}\cdot \hat{v}=0 &\Rightarrow & \sin\theta\cos\phi=0 \label{eqnA8}\\
\hat{e}_{2}\cdot \hat{v}>0 &\Rightarrow &(b_{1}\cos\phi + b_{2}\sin\phi)\sin\theta >0 \label{eqnA9}\\
\hat{e}_{3}\cdot \hat{v}>0 &\Rightarrow &(c_{1}\cos\phi + c_{2}\sin\phi)\sin\theta \nonumber \\
 &~~~~&+ c_{3}\cos\theta >0 \label{eqnA10}
\end{eqnarray}
Now, the inequality (\ref{eqnA9})$\Rightarrow \sin\theta \neq 0$. Therefore, from Eq.(\ref{eqnA8}) one can conclude that $\cos\phi=0 \Rightarrow \phi=\frac{\pi}{2}$ or $-\frac{\pi}{2}$. Next, substituting $\cos\phi=0$ in the inequality (\ref{eqnA9}) one gets $b_{2}\sin\theta\sin\phi>0$. Then, it follows that $\phi=\mbox{sgn}(b_{2})\frac{\pi}{2}$ (since $\sin\theta>0$) thereby reducing the expression for the function $f$ at the interior points of this side of the triangle $R_{+++}$ to
\begin{equation}
f(\theta,\mbox{sgn}(b_{2})\frac{\pi}{2})=f(\theta)=\mbox{sgn}(b_{2})q\sin\theta + r\cos\theta
\end{equation}
Therefore, one gets $\frac{d^{2}f}{d\theta^{2}}=-f(\theta)<0$ which implies that a stationary point in the interior of this side cannot be a point of minimum. Thus, the minimum value  of $f$ can be attained only at some end point(s) of this side.

For the remaining two sides of the triangle $R_{+++}$ on the respective great circles $C_{2}$ and $C_{3}$, similar analyses can be done by choosing the relevant convenient co-ordinate axes. For the side lying on $C_{2}$ ($C_{3}$), the convenient choice is the $X$-axis to be along $\hat{e}_{2}$ ($\hat{e}_{3}$) and the $X-Y$ plane containing $\hat{e}_{3}$ ($\hat{e}_{1}$). Then, it is again found that the minimum value of $f$ on the remaining two sides can only occur at some end point(s) of these sides.    

\section{Maximum violation of the LNR inequality (\ref{eqn6})}
Let us express the LNR inequality (\ref{eqn6}) given in the main text in the following form 
\begin{eqnarray}
S_{AB}(\hat{a}_{1},\hat{a}_{2},\hat{a}_{3},\hat{b}_{1},\hat{b}_{2},\hat{b}_{3},\hat{b'}_{1},\hat{b'}_{2},\hat{b'}_{3})\equiv   \frac{1}{3}\sum _{i}|\langle A_{i}B_{i}\rangle +\nonumber \\ \langle A_{i}B_{i}'\rangle| +\frac{2}{3} \sin (\frac{\beta_{*}}{2}) \times L_{n}(\hat{n}_{1},\hat{n}_{2},\hat{n}_{3})-2 \leq 0\nonumber \\\label{aeqn1} 
\end{eqnarray}
where angles between Bob's settings $\hat{b}_{i}$ and $\hat{b'}_{i}$, $i\in\{1,2,3\}$, are $\beta_{i}$'s with $\beta_{*}=\mbox{min}\{|\beta _{1}|,|\beta _{2}|,|\beta _{3}|\}$ and $\hat{n}_{i}$'s are unit vectors along the directions of $\hat{b}_{i}-\hat{b'}_{i}$. $S_{AB}$ represents a real valued function of settings of Alice and Bob, while the settings for which the quantum mechanically calculated value $S_{AB}>0$ would imply a violation of the LNR inequality given by the preceding inequality (\ref{aeqn1}).

For a pure singlet state, since the QM correlation function $\langle AB\rangle = -\hat{a}\cdot \hat{b}$, the expression (\ref{aeqn1}) reduces to 
\begin{eqnarray}
S_{AB}(\hat{a}_{1},\hat{a}_{2},\hat{a}_{3},\hat{b}_{1},\hat{b}_{2},\hat{b}_{3},\hat{b'}_{1},\hat{b'}_{2},\hat{b'}_{3})\equiv \frac{1}{3}\sum _{i}|\hat{a}_{i}\cdot\nonumber \\(\hat{b}_{i}+\hat{b'}_{i})|+\frac{2}{3} \sin (\frac{\beta_{*}}{2}) \times L_{n}(\hat{n}_{1},\hat{n}_{2},\hat{n}_{3})-2 \leq 0 \nonumber \\\label{aeqn2} 
\end{eqnarray}

From the inequality (\ref{aeqn2}) one can see that as the first step towards maximizing the function $S_{AB}$, Alice's settings $\hat{a}_{i}$'s should lie along the directions of $\hat{b}_{i}+\hat{b'}_{i}$. Then, for maximizing $S_{AB}$, it is sufficient to maximize the function
\begin{equation}
 S_{B}\equiv\frac{1}{3}\sum _{i}|2 \cos\frac{\beta_{i}}{2}|+\frac{2}{3} \sin (\frac{\beta_{*}}{2}) \times L_{n}(\hat{n}_{1},\hat{n}_{2},\hat{n}_{3})-2 \label{aeqn3}
 \end{equation}
 involving only Bob's settings.

Next, note that as the function $L_{n}(\hat{n}_{1},\hat{n}_{2},\hat{n}_{3})$ appearing in the expression (\ref{aeqn3}) does not depend on the values of angles $\beta_{i}$'s, one needs to maximize $L_{n}$ over the directions $\hat{n}_{i}$'s.
 Now, we proceed to show that $\mbox{max}(L_{n}) =1$ when the set $\{\hat{n}_{1},\hat{n}_{2},\hat{n}_{3}\}$ is orthonormal. For this, let us first show that the expression for $L_{n}$ given in a form proved in the theorem of the main text, i.e,
 \begin{equation}
L_n=\frac{|\hat{n}_{1}\cdot (\hat{n}_{2}\times \hat{n}_{3})|}{\mbox{max}\{ |\hat{n}_{1}\times \hat{n}_{2}|, |\hat{n}_{2}\times \hat{n}_{3}|, |\hat{n}_{3}\times \hat{n}_{1}| \}} \label{aeqn4'}
\end{equation}
is equivalent to the expression
 \begin{equation}
L_{n}(\alpha _{12},\alpha _{23},\alpha _{31})=\frac{\Biggl(1\!\!-\!\!\!\displaystyle\sum_{\substack{1\leq i\leq 3, \\j=i\oplus 1}}\!\!\!\cos ^{2}\alpha _{ij} + 2 \!\!\!\!\displaystyle\prod_{\substack{1\leq i\leq 3, \\j=i\oplus 1}}\!\!\!\cos \alpha _{ij}\Biggr)^{\frac{1}{2}}}{\mbox{max}\{\sin \alpha _{12} ,\sin \alpha _{23} ,\sin \alpha _{31}\}} \label{aeqn4}
\end{equation}
which is Eq.(\ref{eqn7}) of the main text. 
The equivalence between (\ref{aeqn4'}) and (\ref{aeqn4}) can be seen as follows. 
Say, $0<\alpha_{12}<\pi$ is the angle between unit vectors $\hat{n}_{1}$ 
and $\hat{n}_{2}$ ($\alpha_{23}$ and $\alpha_{31}$ are defined similarly). For 
convenience, we use the notation $\alpha_{ij}$ for the angle between unit 
vectors $\hat{n}_i$ and $\hat{n}_j$, where $1\leq i \leq 3$ and $j=i\oplus 1$ 
(here $\oplus$ denotes addition modulo $3$).
Then, the denominators of the expressions on the right hand side of the equations (\ref{aeqn4'}) and (\ref{aeqn4}) are same 
since $|\hat{n}_{i}\times \hat{n}_{j}|= |\sin\alpha_{ij}|=\sin\alpha_{ij}$.
Next, we show that the numerators of the two expressions are also equal. 
For this, recall that $\hat{n}_1=(1,0,0)$, $\hat{n}_2=(b_1,b_2,0)$, $\hat{n}_3=(c_1,c_2,c_3)$ (without any loss of generality). 
Then, we find that $|\hat{n}_{1}\cdot (\hat{n}_{2}\times \hat{n}_{3})|=|b_2 c_3|$. Along with this, we also have the following relations $\hat{n}_1\cdot\hat{n}_2=b_1=\cos\alpha_{12}$, $\hat{n}_2\cdot\hat{n}_3=b_1c_1+b_2c_2=\cos\alpha_{23}$, $\hat{n}_3\cdot\hat{n}_1=c_1=\cos\alpha_{31}$  and 
the normalization relations $b_1^{2}+b_2^{2}=1$, $c_1^{2}+c_2^{2}+c_3^{2}=1$. 
By using these relations we find that
$|\hat{n}_{1}\cdot (\hat{n}_{2}\times \hat{n}_{3})|=|b_2 c_3|=
(1-\cos^2\alpha_{12}-\cos^2\alpha_{23}-\cos^2\alpha_{31}+ 2 \cos\alpha_{12}\cos\alpha_{23}\cos\alpha_{31})^{\frac{1}{2}}$

Now, we show that $L_{n}\leq1$ for which note that 
\begin{widetext}
\begin{eqnarray}
-\cos^2\alpha_{23}-\cos^2\alpha_{31}+2\cos\alpha_{12}\cos\alpha_{23}\cos\alpha_{31}
=-\{(\cos\alpha_{23}-\cos\alpha_{31})^{2}+2\cos\alpha_{23}\cos\alpha_{31}(1-\cos\alpha_{12})\}\nonumber\\
=-\{(\cos\alpha_{23}+\cos\alpha_{31})^{2}-2\cos\alpha_{23}\cos\alpha_{31}(1+\cos\alpha_{12})\}\leq 0\nonumber\\
\Rightarrow 1-\cos^2\alpha_{12}-\cos^2\alpha_{23}-\cos^2\alpha_{31}+2\cos\alpha_{12}\cos\alpha_{23}\cos\alpha_{31}\leq \sin^2\alpha_{12}\nonumber\\
\Rightarrow \frac{(1-\cos^2\alpha_{12}-\cos^2\alpha_{23}-\cos^2\alpha_{31}+2\cos\alpha_{12}\cos\alpha_{23}\cos\alpha_{31})^{1/2}}{\sin\alpha_{12}}\leq 1\label{aeqn5}
\end{eqnarray}
\end{widetext}
Similar other inequalities of the form (\ref{aeqn5}) can also be obtained in which $\sin\alpha_{12}$ in the denominator of the left hand side of (\ref{aeqn5}) is replaced by $\sin\alpha_{23}$ or $\sin\alpha_{31}$. Then, combining three such inequalities it can be easily seen that $L_{n}\leq1$ where the maximum value of $L_{n}=1$ occurs, for example, when $\alpha_{12}=\alpha_{23}= \alpha_{31}=\frac{\pi}{2}$.

Therefore, for maximizing $S_{B}$ given by Eq.(\ref{aeqn3}) now it sufficient to maximize the following expression
\begin{equation} 
S_{\beta_{1},\beta_{2},\beta_{3}}\equiv\frac{2}{3}\sum _{i}| \cos\frac{\beta_{i}}{2}|+\frac{2}{3} \sin (\frac{\beta_{*}}{2})-2.\label{aeqn6}
\end{equation}

 Now, note that from Eq.(\ref{aeqn6}) it can be seen that for any given $\beta_{1},\beta_{2},\beta_{3}$ the value of $S_{\beta_{1},\beta_{2},\beta_{3}}$ is bounded by $S_{\beta_{*},\beta_{*},\beta_{*}}$
 \begin{equation}
 S_{\beta_{1},\beta_{2},\beta_{3}}\leq S_{\beta_{*},\beta_{*},\beta_{*}}= 2\cos\frac{\beta_{*}}{2}+\frac{2}{3} \sin \frac{\beta_{*}}{2}-2 
  \end{equation}
Then, we obtain that $\mbox{max}(S_{\beta_{1},\beta_{2},\beta_{3}})\approx0.108$ 
which occurs at $|\beta _{1}|=|\beta _{2}|=|\beta _{3}|=\beta_{*}\approx36.9^{o}$. \hspace{5 mm}$\Box$

Therefore, to summarize, the above derivation implies that the settings of Alice and Bob for maximum QM violation of the LNR inequality (\ref{eqn6}) in the main text are as follows: (i) Alice's settings $\hat{a}_{i}$'s are in the directions of $\hat{b}_{i}-\hat{b'}_{i}$, and (ii) Bob's settings are such that $\hat{n}_{i}$'s are mutually orthogonal and $|\beta _{1}|=|\beta _{2}|=|\beta _{3}|\approx36.9^{o}$. 

\begin{acknowledgments}
We thank V. Scarani, Y. Hasegawa and G. Kar for fruitful discussions. This work is supported by the DST Project SR/S2/PU-16/2007. DH also thanks the Centre for Science, Kolkata for support.
\end{acknowledgments}

\end{document}